# Modeling of Bend Discontinuity in Plasmonic and Spoof Plasmonic Waveguides

Yaghoob Rahbarihagh, Jalil Rashed-Mohassel, and Mahmoud Shahabadi
Center of Excellence on Applied Electromagnetic Systems, School of ECE
College of Engineering, University of Tehran, Tehran, Iran
yrahbari@ut.ac.ir, jrashed@ut.ac.ir, shahabad@ut.ac.ir.

*Abstract*— **The paper proposes a method to characterize the bend discontinuity for plasmonic and spoof plasmonic waveguides in terms of scattering parameters. By means of this method, the waveguide is modelled by a two-port network and its scattering parameters are extracted. The parameters for the L-shaped sharp curved bends at different frequencies and under different bending angles are determined.**

## I. Introduction

Size reduction of optical and electronic devices has always been a technological challenge. But in addition to fabrication constraints, diffraction limit is one of the fundamental restrictions which specifically hinder waveguide size reduction [*1*]. To overcome this limitation, plasmonic waveguides in which surface plasmon polaritons (SPPs) enable wave confinement have been introduced for optical wavelengths. At terahertz frequencies and for submillimeter waves, one alternatively achieves wave confinement using the so-called spoof SPP [2]. The latter uses a periodic structure to realize the reactive boundary condition required for wave confinement.

An array of cylindrical metallic nanorods which can support SPPs have been widely exploited as a waveguide for optical wavelengths. Moreover, their modal analysis has been the subject of several research works [3,4]. The present work concerns with the numerical modeling of bend discontinuities in such waveguides of metallic nanorods. The analysis can be extended to similar discontinuities in spoof SPP waveguides with application at terahertz and submillimeter frequencies.

Characterization of bends in plasmonic waveguides has been the subject of recent researches. Due to increased scattering in the sharp bends of waveguides composed of metallic nanoparticles, an alternative type of plasmonic waveguides is used for wave propagation. Sharp metal wedges and nanogrooves in metal substrates are of special interest for this purpose [5]. The authors investigate propagation of channel plasmon polaritons (CPPs) excited in the wavelength range of 1425–1640 nm along smoothly bent and split V-shaped grooves milled in a gold film. Using channel waveguide with evolution of the channel [6] and implementation in the bending [7] are strategies used in this type of strutures to improve waveguiding. In [8] propagation in the sharp bending in cylindrical waveguides based on silver nanowires is investigated. Power transmission coefficient (η) for different modes are calculated using the dipole approximation propagation, and it is shown that a complete propagation (η=100%) with conversion of transverse and longitudinal modes to each other is possible. In [9] transmission properties of a 90-degree bend in a metallic slot waveguide are studied. In [10], it is shown that bends and splitters with no additional loss over a very wide frequency range can be designed for metal-dielectric-metal plasmonic waveguides.

In [3], propagation of light along an infinite 2D chain of silver nanorods with r=25nm and a separation of L=55nm is analyzed and propagating modes for this waveguide is computed. Here, Generalized Multipole Technique is used for the analysis. In this paper, the behavior of bends in this waveguides is investigated. To this end, the waveguide is modelled by a two-port network and its scattering parameters are extracted. The parameters for the L-shaped sharp bend at different frequencies and with different bending angles were obtained and analyzed.

## II. Modeling

To investigate wave propagation along a waveguide, it is necessary to find the scattering parameters of the waveguide. For doing so, the waveguide can be considered as a two-port network and its scattering parameters are calculated using measured input and output powers.

As shown in [3], the waveguide is capable of supporting odd and even modes in a frequency range of $6.5 \times 10^{14} Hz < f < 7.9 \times 10^{14} Hz$ ($0.12 < L/\lambda < 0.145$ in which L is the particles separation ) (Fig. 1a). Therefore, the waveguide dispersion matrix is a $4 \times 4$ matrix as follows:

$$\begin{bmatrix} b_{1E} \\ b_{1O} \\ b_{2E} \\ b_{2O} \end{bmatrix} = \begin{bmatrix} S_{1E1E} & S_{1E1O} & S_{1E2E} & S_{1E2O} \\ S_{1O1E} & S_{1O1O} & S_{1O2E} & S_{1O2O} \\ S_{2E1E} & S_{2E1O} & S_{2E2E} & S_{2E2O} \\ S_{2O1E} & S_{2O1O} & S_{2O2E} & S_{2O2O} \end{bmatrix} \begin{bmatrix} a_{1E} \\ a_{1O} \\ a_{2E} \\ a_{2O} \end{bmatrix}, \quad (1)$$

where *b* is the output power, *a* is the input power and S is the scattering parameters. Subscripts of parameters *a* and *b* denotes the desired mode and port. The subscript of dispersion scattering parameters is composed of two parts. The first subscript indicates the output mode and the second represents the input mode or source. For example, $S_{1E2O}$ is defined by the following equation:

$$S_{1E2O} = \left.\frac{b_{1E}}{a_{2O}}\right|_{a=0, a \neq a_{2o}}. \quad (2)$$

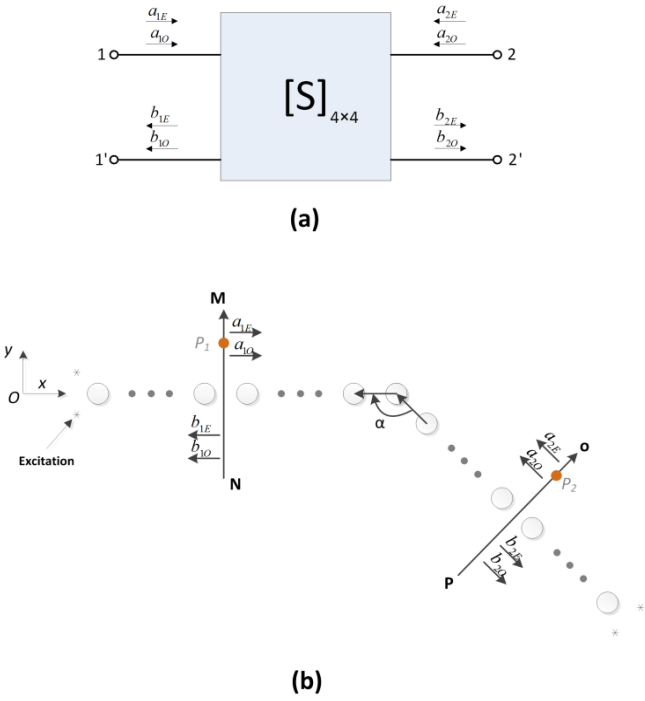

Fig. 1. (a) Waveguide modeling as a two port network and input and output modes. (b) Bend in the waveguide and cross section that are considered to input and output ports.

This means that proportion of output even mode from port 1 to input odd mode from port 2 when the waveguide is excited only by odd mode from port 2 with matched impedances to eliminate other input modes. For waveguides such as rectangular waveguides where power is confined in certain areas, the input and output powers can be easily defined. But for this type of plasmonic waveguides where power is spread in whole space, an accurate definition is needed. Fig. 1b shows the bend. To define input and output ports, MN and OP are defined as port 1 and port 2, respectively. These cross-sections should be far from the bend and excitation points, since reflections from bends have higher modes which can disturb our two mode assumptions. For each input and output mode poynting vector can be written as

$$\begin{aligned}
S_E^+ &= \frac{1}{2} Real\left\{A_E e_E(0,y) \times (A_E h_E(0,y))^*\right\} \\
S_E^- &= \frac{1}{2} Real\left\{-B_E e_E(0,y) \times (B_E h_E(0,y))^*\right\}, \\
S_o^+ &= \frac{1}{2} Real\left\{A_o e_o(0,y) \times (A_o h_o(0,y))^*\right\}, \\
S_o^- &= \frac{1}{2} Real\left\{-B_o e_o(0,y) \times (B_o h_o(0,y))^*\right\},
\end{aligned} \quad (3)$$

where (+) and (-) superscripts indicate forward and backward modes, respectively. Hence, input and output power is obtained as follows:

$$\begin{aligned}
P_E^+ &= \frac{|A_E|^2}{2} \int_{-H}^{H} Real\left\{e_E(0,y) \times (h_E(0,y))^*\right\}.dy \\
&= \frac{|A_E|^2}{2} . PN_E, \\
P_E^- &= -\frac{|B_E|^2}{2} \int_{-H}^{H} Real\left\{e_E(0,y) \times (h_E(0,y))^*\right\}.dy \\
&= -\frac{|B_E|^2}{2} . PN_E, \\
P_o^+ &= \frac{|A_o|^2}{2} \int_{-H}^{H} Real\left\{e_o(0,y) \times (h_o(0,y))^*\right\}.dy \\
&= \frac{|A_o|^2}{2} . PN_o, \\
P_o^- &= -\frac{|B_o|^2}{2} \int_{-H}^{H} Real\left\{e_o(0,y) \times (h_o(0,y))^*\right\}.dy \\
&= -\frac{|B_o|^2}{2} . PN_o.
\end{aligned} \quad (4)$$

where $PN_E$ and $PN_o$ indicate normalized powers of even and odd modes. H is the integral bound and should be selected large enough. According to the definition, *a* and *b* are defined as following:

$$\begin{aligned}
a_{E\atop o} &= \sqrt{PN_{E\atop o}} . A_{E\atop o}, \\
b_{E\atop o} &= \sqrt{PN_{E\atop o}} . B_{E\atop o}.
\end{aligned} \quad (5)$$

To calculate scattering parameters, the normalized power in specific locations such as P1 and P2 in figure1b should be known.

Using the generalized multipole technique discussed in detail in [3], the normalized power for even and odd modes in different frequencies has been obtained (TABLE I). In these calculations, we take H=300r which suffices for our purpose and the normalized point coordinates are x=L/2, y=r with origin at the center of one particle.

TABLE I. POWERS FOR EVEN AND ODD MODES (μW)

| L/λ → Mode ↓ | 0.12 | 0.125 | 0.13 | 0.135 | 0.14 |
|---|---|---|---|---|---|
| Even Mode | 59.7 | 56.8 | 56.9 | 57.4 | 76.6 |
| Odd Mode | 965.5 | 383.5 | 203.6 | 214.9 | 700.3 |

## III. ANALYSIS

To calculate the 4 × 4 scattering matrix with 16 elements, we require 16 measurements or 16 equations. Symmetry can be used to reduce the unknowns here. The structure has symmetry against bisector of α angle in Fig.1b. Therefore we can write the following expressions for the dispersion parameters:

$$S_{1E1E} = S_{2E2E}, \quad S_{1O1E} = S_{2O2E},$$
$$S_{1E1O} = S_{2E2O}, \quad S_{1O1O} = S_{2O2O},$$
$$S_{1E2E} = S_{2E1E}, \quad S_{1O2E} = S_{2O1E}, \quad (6)$$
$$S_{1E2O} = S_{2O1E}, \quad S_{1O2O} = S_{2O1O}.$$

Given these equations, there are only 8 unknown parameters to be calculated.

As mentioned above, to reduce the effects of higher modes, cross-sections should be far enough from the bend and the excitation. In other words, the length of each bend arm should be long enough. The longer length of the waveguide and increased numbers of particles increase the computation time exponentially. Here, we can use a symmetrical excitation with bisector of α and find the field in only one part and reduce the volume and computation time. It should be noted that in this case the numbers of equations are reduced to half. However, since the size and computation time depends exponentially on the number of variables, the use of symmetrical stimulation is effective in reducing the size and computational time.

Using the described method, a bend with a 199-nm particle is analyzed. Each arm consists of 100 2D nanocylinders. The MN and OP cross-sections are selected exactly in the middle of each arm.

Four combinations, even or odd, with respect to the x-axis or the bisector of α, are selected for simulations. Table II shows the scattering parameters for a sharp bend (Fig .1b) with $\theta = 30°$ ($\theta = 180 - \alpha$) at different frequencies.

TABLE II. SCATTERING PARAMETERS FOR A 30° L SHAPE SHARP BEND.

| L/λ→ <br> S ↓ | 0.125 | 0.13 | 0.135 | 0.14 |
|---|---|---|---|---|
| $S_{1E1E}$ | 0.038∠-118° | 0.038∠-15° | 0.038∠57° | 0.031∠53° |
| $S_{1O1E}$ | 0.037∠133° | 0.062∠-115° | 0.046∠-59° | 0.018∠-133° |
| $S_{2E1E}$ | 0.498∠-79° | 0.521∠31° | 0.421∠104° | 0.294∠101° |
| $S_{2O1E}$ | 0.706∠-24° | 0.424∠70° | 0.449∠110° | 0.362∠-42° |
| $S_{1E1O}$ | 0.037∠133° | 0.062∠-115° | 0.046∠-59° | 0.018∠-133° |
| $S_{1O1O}$ | 0.020∠97° | 0.015∠-146° | 0.018∠-98° | 0.009∠133° |
| $S_{2E1O}$ | 0.284∠-63° | 0.368∠67° | 0.168∠115° | 0.054∠35° |
| $S_{2O1O}$ | 0.240∠9° | 0.310∠-88° | 0.190∠-74° | 0.132∠168° |

Input power ($P_{in}$) splits in the waveguide into four parts: output power ($P_{out}$) (which the objective is to increase this part), reflected power due to mismatch ($P_{Mis}$), ohmic loss in the particles ($P_{Loss}$) and radiated power at the bend ($P_{Rad}$). The two latter parts are total power dissipated in the structure. With the scattering matrix, the output power and the power mismatch are:

$$P_{out} = \frac{1}{2}\left(|b_{2E}|^2 + |b_{2O}|^2\right)$$
$$= \frac{1}{2}\left(|S_{2E1E}.a_{1E} + S_{2E1O}.a_{1O}|^2 + |S_{2O1E}.a_{1E} + S_{2O1O}.a_{1O}|^2\right), \quad (7)$$
$$P_{Mis} = \frac{1}{2}\left(|b_{1E}|^2 + |b_{1O}|^2\right)$$
$$= \frac{1}{2}\left(|S_{1E1E}.a_{1E} + S_{1E1O}.a_{1O}|^2 + |S_{1O1E}.a_{1E} + S_{1O1O}.a_{1O}|^2\right).$$

To measure the ohmic loss within the particles we can integrate the poynting vector on the surface of particles. Nevertheless, the summation of the loss and radiation can be calculated from the following equation:

$$P_{Rad} + P_{Loss} = P_{in} - (P_{out} + P_{Mis}). \quad (8)$$

As is clear from the above equations, in addition to the structure, the contribution of each mode in the propagation also depends on excitation. Exciting waveguide with even and odd modes creates different transmission, mismatch, loss and radiation powers. Having said that, the following relationships can be written:

$$\left(\frac{P_{out}}{P_{in}}\right)_{|E} = |S_{2E1E}|^2 + |S_{2O1E}|^2, \quad (9)$$

$$\left(\frac{P_{out}}{P_{in}}\right)_{|O} = |S_{2E1O}|^2 + |S_{2O1O}|^2, \quad (10)$$

$$\left(\frac{P_{Mis}}{P_{in}}\right)_{|E} = |S_{1E1E}|^2 + |S_{1O1E}|^2, \quad (11)$$

$$\left(\frac{P_{out}}{P_{in}}\right)_{|O} = |S_{1E1O}|^2 + |S_{1O1O}|^2, \quad (12)$$

$$\left(\frac{P_{Rad+Loss}}{P_{in}}\right)_{|E} = 1 - \left(|S_{2E1E}|^2 + |S_{2O1E}|^2 + |S_{1E1E}|^2 + |S_{1O1E}|^2\right), \quad (13)$$

$$\left(\frac{P_{Rad+Loss}}{P_{in}}\right)_{|O} = 1 - \left(|S_{2E1O}|^2 + |S_{2O1O}|^2 + |S_{1E1O}|^2 + |S_{1O1O}|^2\right). \quad (14)$$

TABLE III and TABLE IV illustrate the contribution of each power with even and odd excitation. With increasing frequency, less power is transmitted. In other words, we get more ohmic loss and radiation leakage. The high ohmic loss is due to the larger attenuation constant [3]. Furthermore, the odd mode has greater attenuation constant and hence it has more loss.

To investigate the effect of bend angle, scattering parameters have been calculated in the normalized frequency of 0.14. As can be seen form Tables V-VII with increasing the bend angle, mismatch increases. This is reasonable; but it is not valid for other modes and they do not have a linear relation with angle. Since these powers depend on creating higher modes in the bend, little has been known about its mechanism.

TABLE III. CONTRIBUTIONS OF EACH POWER (PERCENT) FOR 30° BEND WITH EVEN EXCITATION.

| L/λ → Power ↓ | 0.125 | 0.13 | 0.135 | 0.14 |
|---|---|---|---|---|
| $P_{out}$ | 76.64 | 45.07 | 37.92 | 21.71 |
| $P_{Mis}$ | 0.93 | 0.89 | 1.21 | 1.44 |
| $P_{Loss} + P_{Rad}$ | 24.43 | 54.04 | 60.87 | 76.85 |

TABLE IV. CONTRIBUTIONS OF EACH POWER (PERCENT) FOR 30° BEND WITH ODD EXCITATION.

| L/λ → Power ↓ | 0.125 | 0.13 | 0.135 | 0.14 |
|---|---|---|---|---|
| $P_{out}$ | 13.85 | 23.11 | 6.44 | 2.04 |
| $P_{Mis}$ | 0.18 | 0.40 | 0.25 | 0.04 |
| $P_{Loss} + P_{Rad}$ | 85.97 | 76.49 | 93.31 | 97.92 |

TABLE V. SCATTERING PARAMETERS FOR L SHAPED SHARP BEND FOR L/Λ=0.14.

| θ → S ↓ | 0° | 30° | 60° | 90° |
|---|---|---|---|---|
| $S_{1E1E}$ | 0.001∡66° | 0.031∡52.7° | 0.137∡55° | 0.177∡85° |
| $S_{1O1E}$ | 0 | 0.116∡-135° | 0.2340∡-119° | 0.201∡-86° |
| $S_{2E1E}$ | 0.439∡99° | 0.294∡101° | 0.094∡91° | 0.161∡-13° |
| $S_{2O1E}$ | 0 | 0.362∡42.5° | 0.365∡31° | 0.075∡-49° |
| $S_{1E1O}$ | 0.037∡133° | 0.018∡-133° | 0.038∡-117° | 0.028∡-92° |
| $S_{1O1O}$ | 0.002∡25° | 0.009∡133° | 0.046∡141° | 0.060∡160° |
| $S_{2E1O}$ | 0 | 0.054∡35° | 0.053∡11° | 0.033∡-95° |
| $S_{2O1O}$ | 0.146∡174° | 0.132∡168° | 0.117∡135° | 0.161∡83° |

TABLE VI. PROPORTION OF EACH POWER (PERCENT) FOR A SHARP BEND IN L/Λ=0.14 WITH EVEN EXCITATION.

| θ → Power ↓ | 0° | 30° | 60° | 90° |
|---|---|---|---|---|
| $P_{out}$ | 19.29 | 21.71 | 14.20 | 3.15 |
| $P_{Mis}$ | 0 | 1.44 | 7.63 | 7.16 |
| $P_{Loss} + P_{Rad}$ | 80.71 | 76.85 | 78.16 | 89.69 |

IV. CONCLUSION

In this paper, we modelled the bends in a plasmonic waveguide based on 2D nanocylindrical particles with a two-port network and using generalized multipole technique. We calculated scattering parameters and investigated the effects of bend angle and frequency on wave propagation.

TABLE VII. CONTRIBUTIONS OF EACH POWER (PERCENT) FOR A SHARP BEND IN L/Λ=0.14 WITH ODD EXCITATION.

| θ → Power ↓ | 0° | 30° | 60° | 90° |
|---|---|---|---|---|
| $P_{out}$ | 2.12 | 2.04 | 1.67 | 2.71 |
| $P_{Mis}$ | 0 | 0.04 | 0.36 | 0.44 |
| $P_{Loss} + P_{Rad}$ | 97.87 | 97.92 | 97.98 | 96.84 |